# Band structure of a two dimensional metallic photonic crystal and the experimental observation of negative refraction in the microwave region

R. A. Gutiérrez-Arenas and D. Mendoza*

*Instituto de Investigaciones en Materiales, Universidad Nacional Autónoma de México*

*Apartado postal 70-360, 04510 México D.F. México*

In this paper we describe an application of the finite difference method to obtain the transverse magnetic photonic band gap diagram of a metallic photonic crystal. The strategy of this method is to formulate the Maxwell equations in finite differences in order to write a computational code. We present experiments that confirm the validity of the calculations of the photonic band diagram as well as the refraction index of such structure. All the calculations were made for a two dimensional metallic photonic crystal.

En este trabajo se presenta la aplicación del método de diferencias finitas para obtener el diagrama de las bandas transversal magnético de un cristal fotónico metálico. El método consiste en formular las ecuaciones de Maxwell en diferencias finitas para poder escribir un código computacional. Además se presentan experimentos que confirman la validez de los cálculos del diagrama de bandas, así como del índice de refracción de la estructura en cuestión. Todos los cálculos y mediciones fueron hechos para un cristal fotónico metálico bidimensional.

## 1. Introduction

Photonic crystals (PC) are periodic dielectric or metallic structures in which the periodicity gives rise to bands for the propagation of electromagnetic (EM) waves and to band gaps where propagation is prohibited for certain frequency range. Using different materials (i.e. with different dielectric constants) and by adjusting the geometrical parameters of the crystal, the band structure can be modified virtually in any way in a controllable manner. The scale invariant nature of the Maxwell's equations enables the study of EM phenomena without being held back by structural complexities [1].

The enormous possibilities for applications in telecommunications and the optical wavelength scale drive the research of PC towards the micro and nanoscales. Also PC present a novel optical phenomenon, namely, negative refraction which produces many applications such as sub-wavelength focusing [2], as well as auto-focusing through a two dimensional (2-D) slab [3,4].

In 1960, Veselago proposed the electrodynamics in a medium possessing a negative index of refraction ($n$) [5]. The realization of such medium has become a subject of interest recently using artificial composite structures with negative dielectric permittivity ($\varepsilon$) and negative magnetic permeability ($\mu$). This phenomenon may lead to novel lens structures that will surpass the limitations of conventional (positive refraction) lenses. For a PC the EM phenomena depend only in the refractive index of the dielectric or metallic material used to build the PC and on the geometrical parameters. Hence, their scalability across the entire EM spectrum makes them potentially available for a wide range of applications. Based on this fact, photonic crystal structures are anticipated to be an essential component of photonic integrated circuits in the near future. Recent studies show an even more *shocking* application of the negative refractive index. Smith *et al.* have suggested that metamaterials (materials with $\varepsilon<0$, and $\mu<0$, and therefore $n<0$) could produce a *cloak of invisibility*. This approach has been reported both theoretical and experimental over a narrow band of microwave frequencies [6,7].

The first step towards the design of photonic crystals is the complete understanding of their physical properties. The band diagram of a PC gives complete information about their optical properties. There are several methods to obtain the photonic band diagram of a photonic crystal: finite difference (in time and coordinate space domain), plane wave expansion and transfer matrix methods.

In this paper, we present the results of the numerical investigation of the wave propagation in a square and triangular lattice of a 2-D metallic PC using the finite difference method. We present only the band diagram for transverse magnetic modes. Finally, comparisons are made between the simulations and recent experiments.

**2. Maxwell's Equations in a discrete mesh**

A 2-D PC is periodic in two of its spatial axis and homogenous in the third. The photonic band gaps appear on the plane of periodicity and also the EM waves propagate in such plane. It is possible to describe completely a metallic PC by its conductivity function or profile ($\sigma$) given in the periodic plane. In this paper we will consider such plane as the one formed by the $x$ and $y$ axis and we will define $\mathbf{x}_{//} = x\hat{\mathbf{x}} + y\hat{\mathbf{y}}$ as the transversal displacement vector. This vector is periodic in all the photonic crystal. Given the fact that the PC has discrete translational symmetry of the conductivity profile, we only need to solve the Maxwell's equations in the fundamental unit cell of the crystal.

It can be shown from Maxwell's equations and from direct symmetry of the PC that the EM waves can be decomposed into two independent modes of propagation: the Transverse Electric (TE) modes and the Transverse Magnetic (TM) modes. In a TE mode the electric field vector is perpendicular to the post axis ($z$-axis) and in a TM mode the magnetic field vector is perpendicular to the post axis. All the field components in the TM or TE modes can be expressed through the axial ($z$) component of the electric or magnetic field [1].

There are essentially two types of arrays for a 2-D PC, the first consists of an array of rods in a dielectric background and the second consists of an array of holes in a dielectric slab. We will center our attention on the former given the fact that the EM modes preferred by this PC are the TM ones [8], and those modes are easier to measure in a controlled laboratory experiment. Moreover this kind of PC can be arranged into two different lattices of rods; square (Fig. 1a) or triangular (Fig. 1b) lattices can be formed to satisfy the 2D periodicity.

If we consider that the rods in the PC are metals and in addition that they are ideal, the conductivity function will have an infinity value in the entire surface and inside the metallic rods. The dielectric background chosen for this PC was the vacuum, with a dielectric constant ($\varepsilon_r$) equal to one.

Taking into account all of these considerations, Maxwell's equations can yield to the Helmholtz equation for the electric field ($\mathbf{E}_{//}$) within the fundamental unit cell (these cells are shown in Fig. 1 depicted by the shaded zone of the diagrams):

$$\nabla_{//}^{2}\mathbf{E}(\mathbf{x}_{//}) = \left(k_z^2 - \frac{\omega^2}{c^2}\right)\mathbf{E}(\mathbf{x}_{//}), \tag{1}$$

where $\omega$ is the angular frequency, $k_z$ is the longitudinal wave number and $c$ is the velocity of the light in the vacuum. The boundary condition on the surfaces $S$ of the conducting posts is:

$$\mathbf{E}\big|_S = 0\,. \tag{2}$$

The discrete translational symmetry of the conductivity profile allows the fundamental solution of the Helmholtz equation to be written in Bloch form and the following periodic conditions are deduced for each lattice:

$$E(a,y) = E(0,y)e^{ik_x a}$$
$$E(x,a) = E(x,0)e^{ik_y a} \quad \text{(square lattice),} \tag{3}$$

$$E\left(a + \frac{1}{\sqrt{3}}y, y\right) = E\left(\frac{1}{\sqrt{3}}y, y\right)e^{ik_x a}$$
$$E\left(x + \frac{a}{2}, \frac{\sqrt{3}}{2}a\right) = E(x,0)e^{ik_x (a/2) + ik_y \sqrt{3}(a/2)} \quad \text{(triangular lattice).} \tag{4}$$

The Helmholtz equation and the boundary conditions described in the former equations define the Eigenvalue problem of finding $k_z^2\text{-}\omega^2/c^2$ as a function of the transversal wave number, $\mathbf{k}_{//} = k_x\hat{\mathbf{x}} + k_y\hat{\mathbf{y}}$ which is also periodic and can be restricted to the contours of the irreducible Brillouin zone of the correspondent reciprocal lattice (shown in Fig. 2 for the square and triangular lattices). For each lattice there exist 3 high symmetry points marked in figure 2 as $\Gamma$, $M$ and $X$.

The FD method solves differential equations by dividing the spatial domain into small partitions. An expression in FD for the second derivative of the function $U(x)$ in the point $x_0$ is given by:

$$U''(x_0) = \left(\frac{d^2U}{dx^2}\right)_{x=x_0} \approx \frac{1}{h^2}\left[U(x_0 + h) - 2U(x_0) + U(x_0 - h)\right] \tag{5}$$

where $h$ is a small and finite value. In this expression the differential equation transforms into an algebraic equation. The derivative is defined by the values of the function in discrete points next to $x_0$.

The strategy of the FD method consists on an approximation of the second order derivative made by a system of $n$ linear equations, where $n$ is an integer and correspond to the number of points used to represent the space where the function $U$ is of interest. Its value and accuracy depends on the value of $h$.

In a two dimensional case, the FD method changes the continuous values of spatial axis $x$ and $y$ to discrete values indexed by integers, $i$ and $j$ respectively, in a way that the spatial values are obtained by the equations $x=ih$ and $y=jh$. For convergence reasons we cover the fundamental unit cell of the square and triangular lattice by a square or triangular mesh respectively with $(2N+1)^2$ mesh points, where $N$ is an integer, i.e. $h$ is the same for the two axis. In both meshes the mesh points are equidistant and follow the geometry of the fundamental unit cell.

Outside the conducting posts, the Helmholtz equation is approximated by the system of linear algebraic equations formed by the values of the electric field in all the mesh points $(i,j)$. For each lattice the discrete Helmholtz equation are the following

$$E_{i+1,j} + E_{i-1,j} + E_{i,j+1} + E_{i,j-1} - 4E_{i,j} = h^2\left(k_z^2 - \omega^2/c^2\right)E_{i,j} \text{ (square lattice)}, \quad (6)$$

$$\begin{aligned} &4\left(E_{i+1,j} + E_{i-1,j} + E_{i,j+1} + E_{i,j-1}\right) + \\ &\left(E_{i+1,j-1} - E_{i+1,j+1} + E_{i-1,j+1} - E_{i-1,j-1}\right) - 16E_{i,j} = 3\left(k_z^2 - \omega^2/c^2\right)E_{i,j} \end{aligned} \text{ (triangular lattice).} \quad (7)$$

To simplify the notation we have used the sub index to indicate the point where is considered the electric field. For example $E_{i+1,j}=E(hi+h,jh)$. It is worth mentioning that the expression referring to the square lattice was obtained by a simple substitution of equation (5) in equation (1), meanwhile for the triangular lattice the fact that the fundamental unit vectors of the fundamental unit cell are not orthogonal an axis transformation was necessary, which yielded to equation (7) [9, 10]. In the FD method $h$ is known as mesh step and it is equal to $a/(2N+1)$. Any value of $k_z$ is valid for both equations, therefore if we select $k_z=0$, the numerical results are valid without any loss of generality.

The boundary conditions in equations (3) and (4) are expressed explicitly as

$$\begin{aligned} E_{2N+1,j} &= E_{0,j}e^{ik_x a} \\ E_{i,2N+1} &= E_{i,0}e^{ik_y a} \end{aligned} \quad \text{(square lattice),} \qquad (8)$$

$$\begin{aligned} E_{2N+1,j} &= E_{0,j}e^{ik_x a} \\ E_{i,2N+1} &= E_{i,0}e^{i(a/2)(k_x+\sqrt{3}k_y)} \end{aligned} \quad \text{(triangular lattice).} \qquad (9)$$

And the mesh points that fall inside the conducting posts are excluded from the system of linear equations given the fact that the value of the electric field inside the post is equal to zero. Thus equation (1) is completely represented by a closed set of $(2N+1)^2-M$ linear equations, where $M$ is the number of points that fall inside the conducting cylinder. If we don't take into account the electrodynamic losses inside the photonic crystal the coefficient matrix of this system is Hermitian [11, 12] and the eigenvalues of such matrix are real. Therefore the problem of obtaining the band diagram of a photonic crystal is reduced to find the eigenvalues of the coefficient matrix and comparing them to the term on the right side of equations (6) and (7) to obtain the values of the allowed frequencies for a given transversal wave vector. The dependence of the eigenvalues of the matrix to the wave vector is given in the boundary conditions expressed in equations (8) and (9). The finite difference method is an iterative process; this method could be presented easily in a concise and clear form using a pseudo code or an algorithm [13].

## 3. The calculation of the refraction index

The research of photonic crystals has been directed to the study of its properties as materials that do not allow the propagation of light in certain frequencies. Nevertheless, PC could be used as light conductors and its conductance is given by the physical structure of the photonic crystal. The refraction index is a property of great importance given the wide range of applications it could have [3, 4, 6-8]. It has been shown that photonic crystals are capable to show a refraction index smaller than unity and even negative and Snell law could be used to describe light propagation in it.

There are two ways to define a refraction index in a PC; Joannopoulos *et al.* [1] obtains the index upon the curvature of the band structure and the second one consists on the form of the band structure near the band gap and it was presented first by Notomi [8]. In this paper, we obtain the refraction index using the ideas posted by Notomi.

For a plane wave with frequency $\omega$, incident on a photonic crystal—air interface, the wave vector inside the crystal $\mathbf{k}_f$, is parallel or antiparallel to the group velocity vector inside the crystal, $\mathbf{v}_g$, depending on the band diagram. If both vectors are parallel, the electromagnetic fields inside the crystals are right handed or ordinary, but if the wave vectors are antiparallel the EM fields are left handed and a negative refractive index could be defined. Therefore inside a photonic crystal the effective refraction index, $n_{eff}$, is defined by:

$$n_{eff} = \mathrm{sgn}\left(\mathbf{v}_g \cdot \mathbf{k}_f\right)\left(\frac{c}{\omega}\left|\mathbf{k}_f\right|\right), \tag{10}$$

where $c$ is the speed of light in vacuum and $\mathbf{v}_g$ is the 2group velocity inside the crystal. The group velocity could be calculated by the general expression $\mathbf{v}_g = \nabla_{\mathbf{k}}\omega$ and using the band diagram of the PC [4].

**4. Results of the finite difference method**

The calculations of the band diagram were made for the two different types of 2-D PC described in this document (square and triangular lattices). We consider ideal metallic rods in a vacuum background with a lattice constant $a$. The radii of the metallic rods were chosen to satisfy the relation $r$=0.2$a$, corresponding to a filling fraction of barely 13% of fundamental unit cell. The diagrams presented in this section are normalized with respect to $a$. for all the graphs, $k_z$=0, which does not affect the generality of the results. The algorithm showed no changes in the results as $N$>20, therefore for both lattices we used 1681 mesh points to represent completely the fundamental unit cell.

**Square lattice**

In figure 3a we show the band diagram for the TM modes of a 2-D PC made out of a square lattice of metallic rods, as the wave vector $\mathbf{k}_{//}$, vary from the far edge of the irreducible Brillouin zone (point $M$ in figure 2a) to the nearest edge ($X$ in figure 2a) and finally to the center of the Brillouin zone ($\Gamma$ in figure 2a). In the graph we can find a band gap between the first and second band ($\omega_1$ and $\omega_2$ respectively). The band gap has an amplitude of $\omega = 0.133(2\pi c/a)$, moreover the system has a cutoff frequency at $\omega = 0.527(2\pi c/a)$.

For most of the photonic crystals structures, the behavior of the effective refractive index can be observed with great ease in the first bands, given the fact that there is less disorder in such bands [14]. The first two bands of the PC analyzed in this section will be studied with the purpose to find a region where we can define an effective refractive index (positive or negative).

The equifrequency surfaces (EFS) representation of the band diagram is an alternate depiction of the photonic band diagram. It consists on the representation of the contours of equal frequency in the first Brillouin zone. There exists an EFS diagram for each photonic band. For a conventional dielectric the EFS diagram are circles centered around the origin of the first Brillouin zone and therefore a refractive index could be defined easily, meanwhile for a photonic crystal the EFS diagrams could have a wide variety of figures and according to [8] the refractive index can only be defined when the contours of equal frequency behave like a dielectric.

The EFS diagrams of the first two bands of the square lattice PC are shown in figure 4. For the first band an effective refractive index can be defined, with values between 0 and 0.4 for a bandwidth of $0.055(2\pi c/a)$. Such index is positive due to the parallel nature of the group velocity vector ($\mathbf{v}_g$) and the wave vector ($\mathbf{k}_{//}$). For band $\omega_2$ the refractive index has values from -0.2 to 0 and a bandwidth of $0.040(2\pi c/a)$. In this band the group velocity vector points toward the center of the Brillouin zone, therefore the index is negative. In figure 5a we show a graphical representation of the refractive index against frequency.

**Triangular lattice**

The first seven bands for the TM modes of a 2-D metallic PC following a triangular lattice are shown in figure 3b. Such diagram is presented modifying the wave vector ($\mathbf{k}_{//}$) from the farthest edge of the ($X$) to the nearest edge ($M$) of the irreducible Brillouin zone and finally to the center ($\Gamma$) of the irreducible Brillouin zone. In the figure we can detect that there is a band gap between bands $\omega_2$ and $\omega_3$ with amplitude (bandwidth) equal to $0.033(2\pi c/a)$, and the cutoff frequency of the system due to the symmetry of the crystal is $0.628(2\pi c/a)$.

In the same way as the square lattice, only the first two bands of the triangular lattice will be analyzed with the purpose to define an effective refractive index. In figure 6 we can examine the EFS representation of bands $\omega_1$, $\omega_2$, and $\omega_3$. For the first to two bands an unambiguous region is found where the notion of an effective refractive index is valid, due to the fact that the PC behaves like a conventional dielectric. On the contrary for $\omega_3$ (figure 7c), the effective refractive index is impossible to define and therefore the transmission of EM waves in this region will be similar to a diffraction grating [8].

The EFS of the first band has a bandwidth of 0.085($2\pi c/a$) where the contours behave like a circle centered on the origin of the Brillouin zone. Also, the group velocity vector is parallel to the wave vector, and therefore the refractive index will be positive. The values of the effective refractive index go from 0 to 0.58. For the second band the EFS are circular in a bandwidth equal to 0.218($2\pi c/a$), and the refractive index is negative given the fact that the group velocity vector is antiparallel to the wave vector and it is pointing towards the center of the Brillouin zone. The values of such index go from -0.45 to 0. A graph depicting these results, for both bands, is presented in figure 5b.

From the information given by the finite difference algorithm, we can conclude that the triangular lattice has a bigger bandwidth where it is possible to watch a region with a well defined refractive index. The triangular lattice presents a region almost five times bigger than the square lattice where we can observe the negative refractive index, and therefore the triangular lattice is better for the observation of the negative refraction phenomena.

**6. Experimental verification of the finite differences algorithm**

A triangular lattice 2-D metallic PC was built to achieve the experimental verification of the finite difference method as well as to show the existence of the negative refraction using the property of auto-focus [15]. The property of auto-focusing is a conclusive effect of the existence of a negative index of refraction. Figure 7a shows an image of the built photonic crystal and 7b shows the experimental scheme used to determine the negative index and the auto-focus.

The physical dimensions of the PC where such that negative refractive index could be observed in the microwave region, more specifically in the X band (8-12 GHz). The materials used in the construction where copper for the metallic rods and they were sandwiched in a couple of brass plates to produce the effect of a parallel plate waveguide. The lattice has the post radius $r$=6.35mm and the distance between the nearest post $a$=31.5mm, which corresponds to $r/a$=0.2. The complete PC was a rectangular slab with dimensions equal to, in lattice parameters units, 7$a$ in width and 18$a$ in length.

Two experiments were developed to fulfill the objectives of this section. The first one consisted on watching the frequency response of the PC from 7 to 13 GHz using a couple of open waveguide antennae, and illuminating the photonic structure in the $\Gamma \rightarrow M$ direction. For this experiment we observed a band gap between 10.97 and 11.48 GHz which is equivalent to 0.51 GHz bandwidth and agrees completely with the calculated numbers thrown in by the finite difference algorithm. The results are shown in figure 8.

The second experiment consisted on an array engineered to measure the negative refractive index. It consisted first, on illuminating one side of the rectangular PC with an open waveguide, fixed at the origin of the *xy* plane showed in figure 7b. Secondly a mapping of the opposite side of the crystal, was made with another open waveguide (this one was situated in the *x* direction at $L+d+s'$ with $s'$ varying from 0 to 35 cm, and in the *y* direction from one edge of the photonic structure to the other, as showed in figure 7b). All measurements were made on the *xy* plane. Finally we could observe a concentration of energy defining in this way an experimental effective refractive index. In figure 9a it is shown a typical mapping of the transmitted energy after crossing the PC structure.

The experimental value of the index of refraction was estimated using Snell's law and the ray diagram shown in figure 7b and it is given by the expression:

$$s' = (L - s)\sqrt{\frac{1-\sin^2(\phi)}{|n_{eff}|^2 - \sin^2(\phi)}}\,, \tag{10}$$

where *s'* is the position of the external focus, $\phi$ is the incident angle, *L* is the width of the photonic crystal and *d* is the distance from the source to the crystal. It is worth mentioning that the position of the external focus is a function of the value of the incident angle, except in the case where $n_{eff}$=-1, therefore for this particular photonic crystal, the auto-focusing region is not punctual. As a rule for defining the values of the negative index for a single frequency, was to obtain the value of *s'* where the power received by the antenna was maximum. And subsequently we obtained the values of *s'*, where the power was half of the maximum. Finally the region of negative index was delimited by those points (see error bars in figure 9b). In figure 9b we present a summary of the experimentally obtained refractive index, which fairly agree with the ones previously calculated. We believe that the observed difference between the calculated and the experimental values of the refractive index may be related to the idealization of the calculated structure. For example, the calculated photonic crystal is infinite and the rods are considered ideal metals (infinite conductivity), meanwhile the experimental crystal is a finite structure and real materials were used. Further study on the optimization of the experimental configuration is necessary.

**7. Conclusions**

In this document we presented an implementation of the finite difference algorithm to calculate the band structure of two different lattices of a two dimensional metallic photonic crystal. This method is very illustrative and helps understand better the theoretical foundations of the photonic crystals. It is also a quite simple algorithm given the fact it could be implemented in several programming platforms, from Matlab to Fortran or C++. We also showed the capabilities of this method and a wide range of applications it offers. The inherent property of photonic crystals of having a range of frequencies where it shows a negative refractive index was also explored in this paper. The band diagrams and the refractive index calculation were verified with two experiments. It is also the purpose of this document to encourage the research on photonic crystals as left handed materials.

## Acknowledgments

The authors wish to thank the Departamento de Electromagnetismo Aplicado at the Facultad de Ingeniería (UNAM) for the help and support in the experimental measurements, in particular Professor M. Ibarra and Dr. O. Martyniuk. Also, the authors would like to thank Dr. M. Ley Koo at the Facultad de Ciencias (UNAM) for allowing us to use the semi-anechoic chamber for the experimental measurements.

## References

* e-mail: doroteo@servidor.unam.mx

1. J. D. Joannopoulos, R.D. Meade, and J. N. Winn, *Photonic Crystals: Molding the flow of light* (Princeton University Press, Boston, 1995).

2. J. B. Pendry, Phys. Rev. Lett. **85**, 3966 (2000).

3. E. Cubukcu, K. Aydin, E. Ozbay, S. Foteinopoulou, and C. M. Soukoulis, Nature **423**, 604 (2003).

4. P. V. Parimi, W. T. LU, P. Vodo, J. Sokoloff, J. S. Dero, and S. Sridhar, Phys. Rev. Lett. **92**, 127401 (2004).

5. V. G. Veselago, Sov. Phys. Usp. **10**, 509 (1968).

6. D. Schurig, J. J. Mock, B. J. Justice, S. A. Cummer, J. B. Pendry, A. F. Starr, and D. R. Smith, Science **304**, 977 (2006)

7. S. A. Cummer, B.I. Popa, D. Schurig, D. R. Smith, and J. B. Pendry, Phys. Rev. E **74**, 036621 (2006).

8. M. Notomi, Phys. Rev. B **62**, 10696 (2000).

9. G. D. Smith, *Numerical solution of partial differential equations: finite difference methods* (Oxford, GB, 1985).

10. C. F. Gerald, *Numerical Analysis* (Adiison—Wesley, NY, 1991).

11. D. R. Smith, S. Schultz, N. Kroll, M. Sigalas, K. M. Ho, and C. M. Soukoulis, Appl. Phys. Lett. **65**, 645 (1994).

12. E. I. Smirnova, C. Chen, M. A. Shapiro, J. R. Sirigiri and R. J. Temkin, J. Appl Phys. **91**, 960 (2002).

13. I) Parameter definition (square or triangular lattice, *r/a*, *N* and the values of $\mathbf{k}_{//}$).

II) Main iteration over the values of the wave vector.

  a) Division of the fundamental unit cell according to the value of mesh points.

  b) Identify the mesh points that fall inside a metallic rod.

  c) Build the coefficient matrix by assigning to each mesh point the Helmholtz discrete equation and the corresponding boundary conditions.

  d) Calculation of the eigenvalues of the coefficient matrix.

  e) Sort the values of the frequencies and save them into the memory.

  f) End of the main iteration

III) Graphing

IV) End of the program.

14. D. R. Smith and N. Kroll, Phys. Rev. Lett. **85**, 2933 (2000).

15. A. Martínez, H. Míguez, A. Griol, and J. Martí, Phys. Rev. B **69**, 165119 (2004).

**Figure 1**

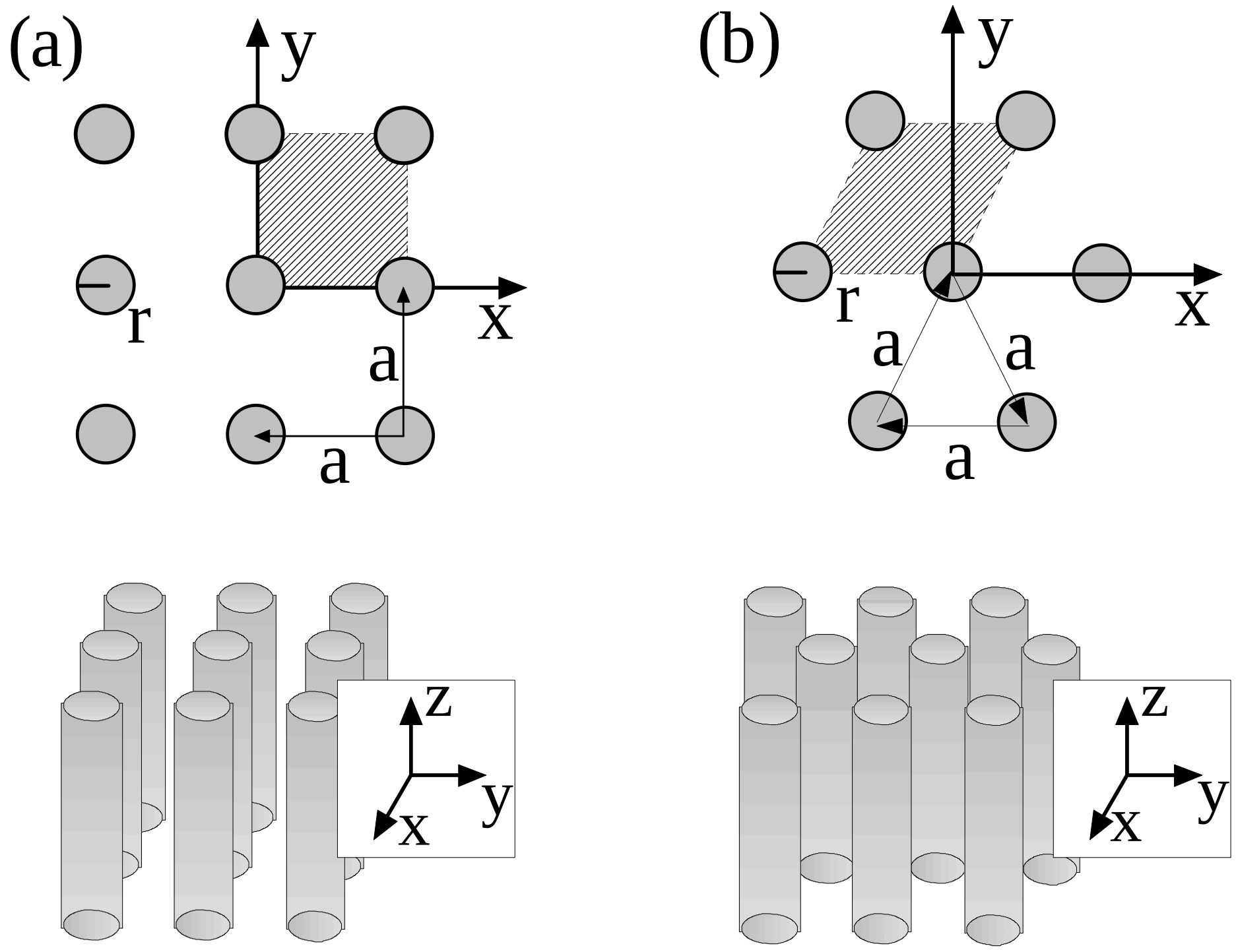

**Figure 2**

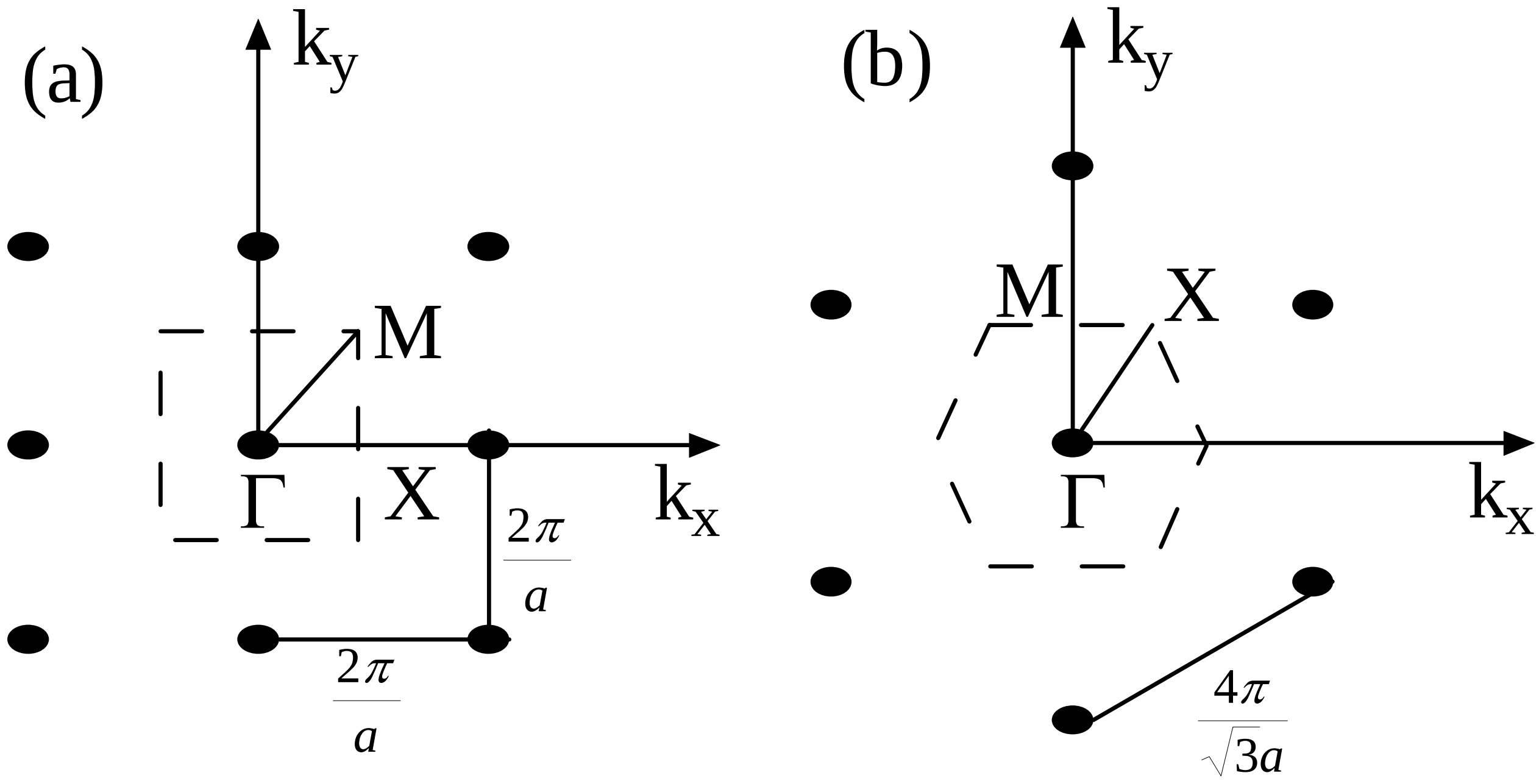

**Figure 3a**

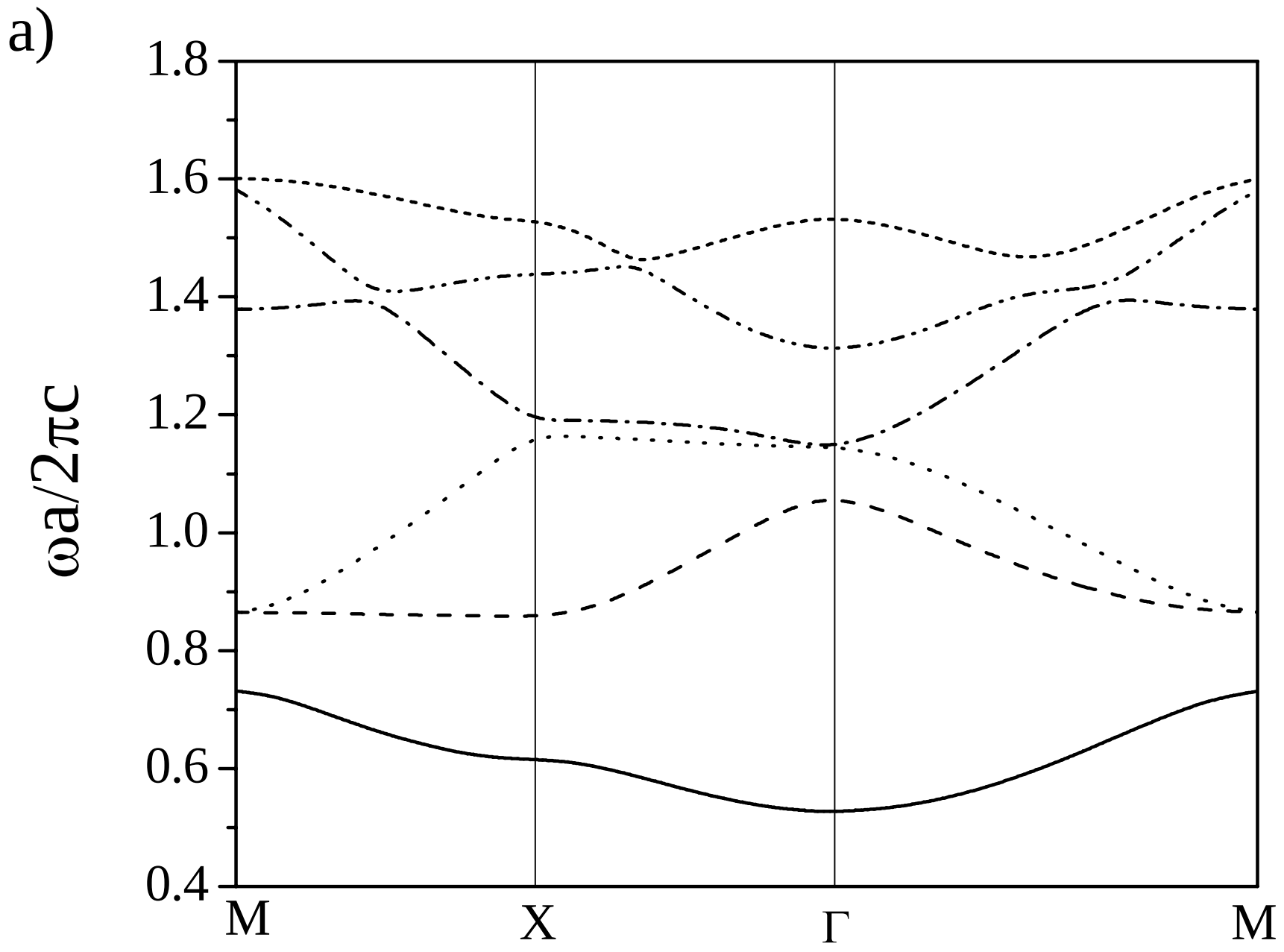

**Figure 3b**

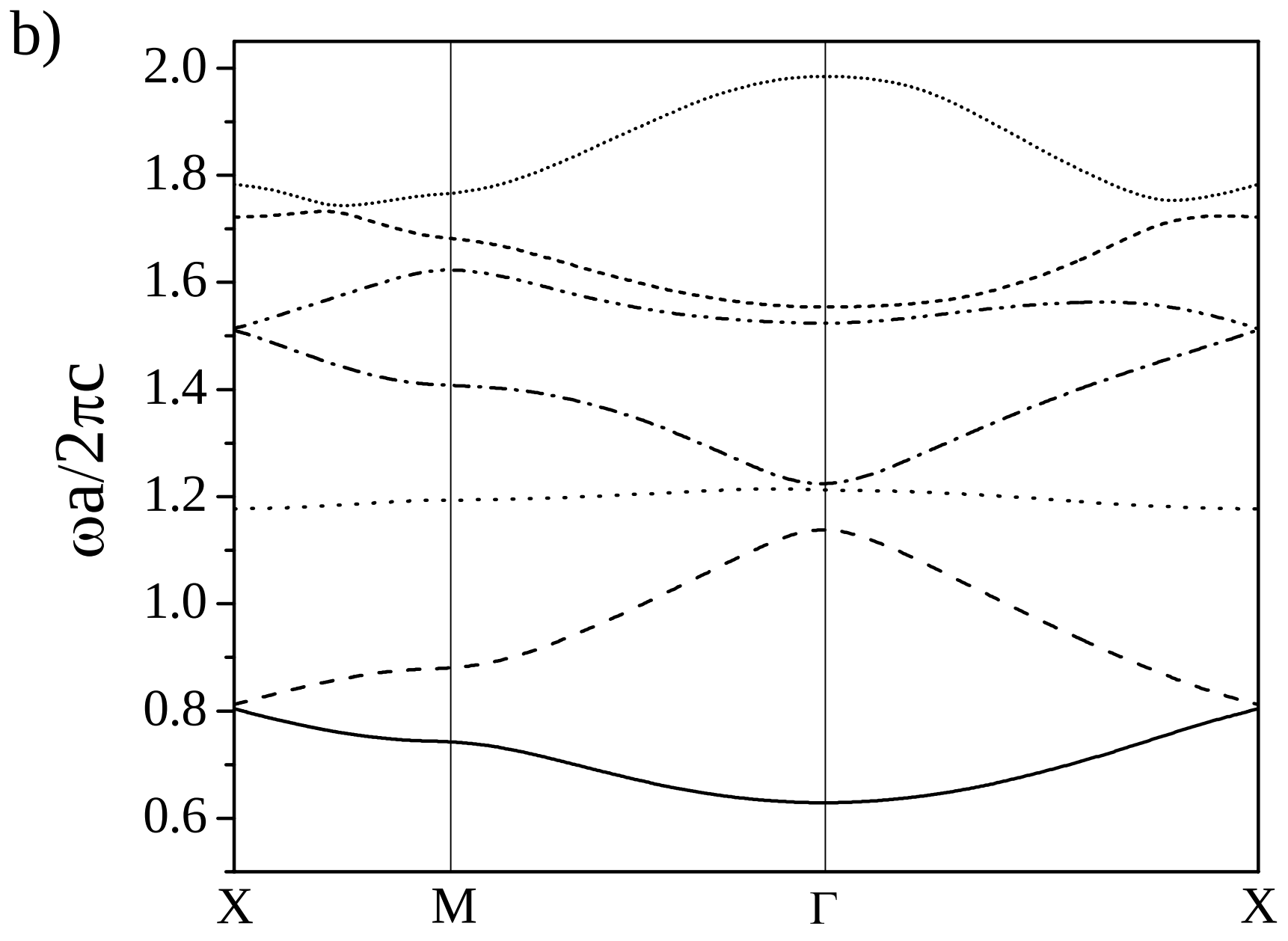

**Figure 4a**

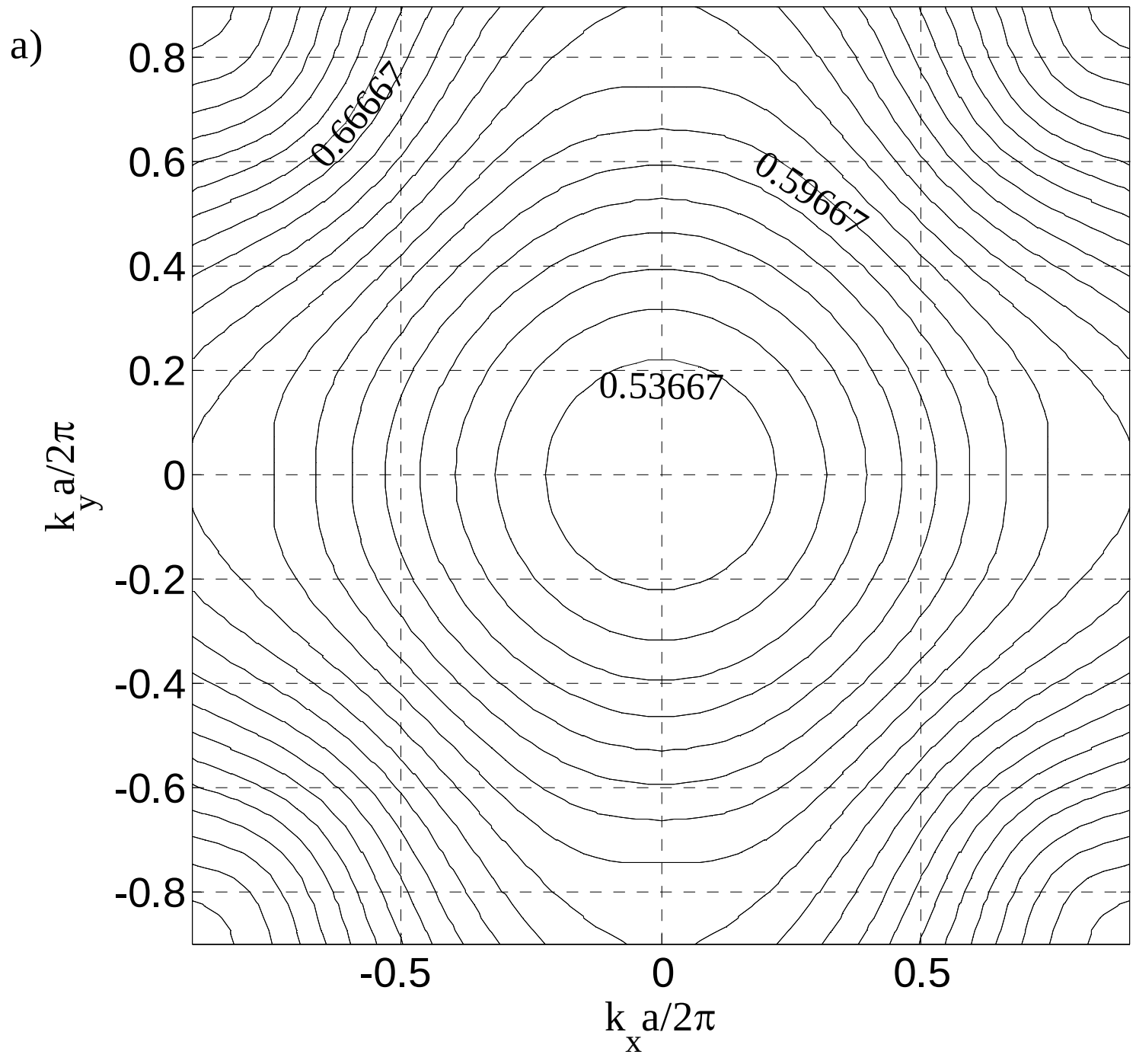

**Figure 4b**

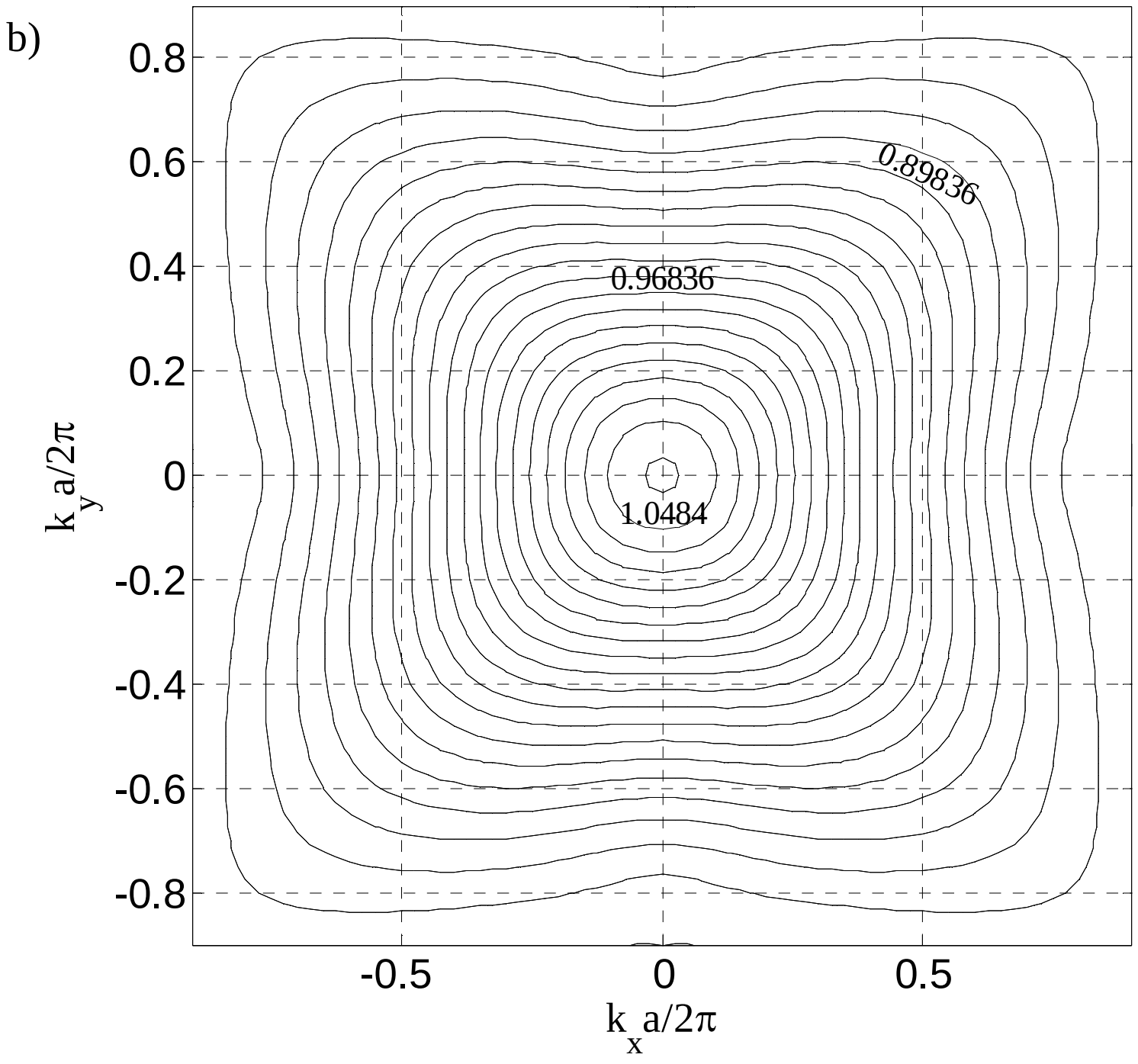

**Figure 5a**

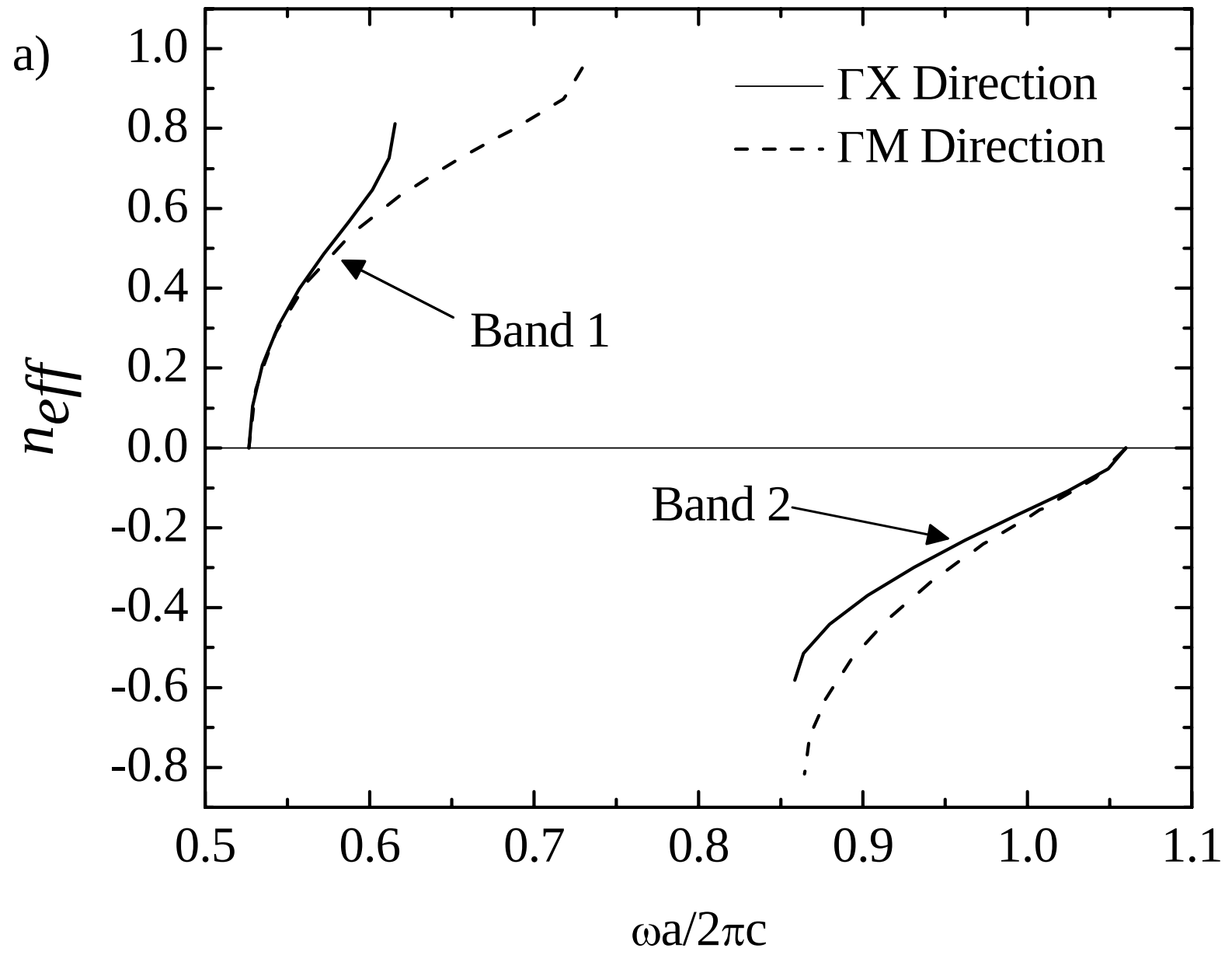

**Figure 5b**

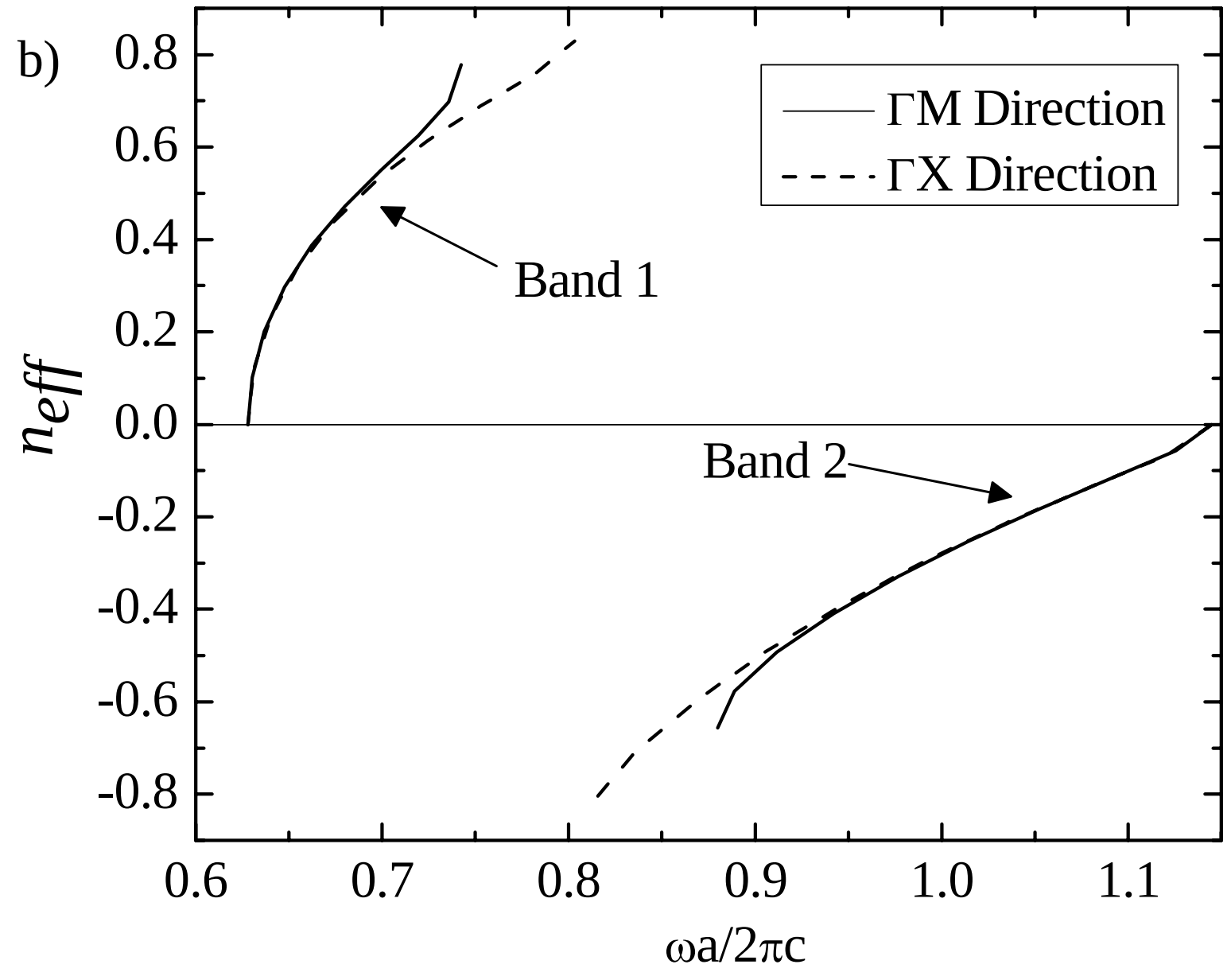

**Figure 6a**

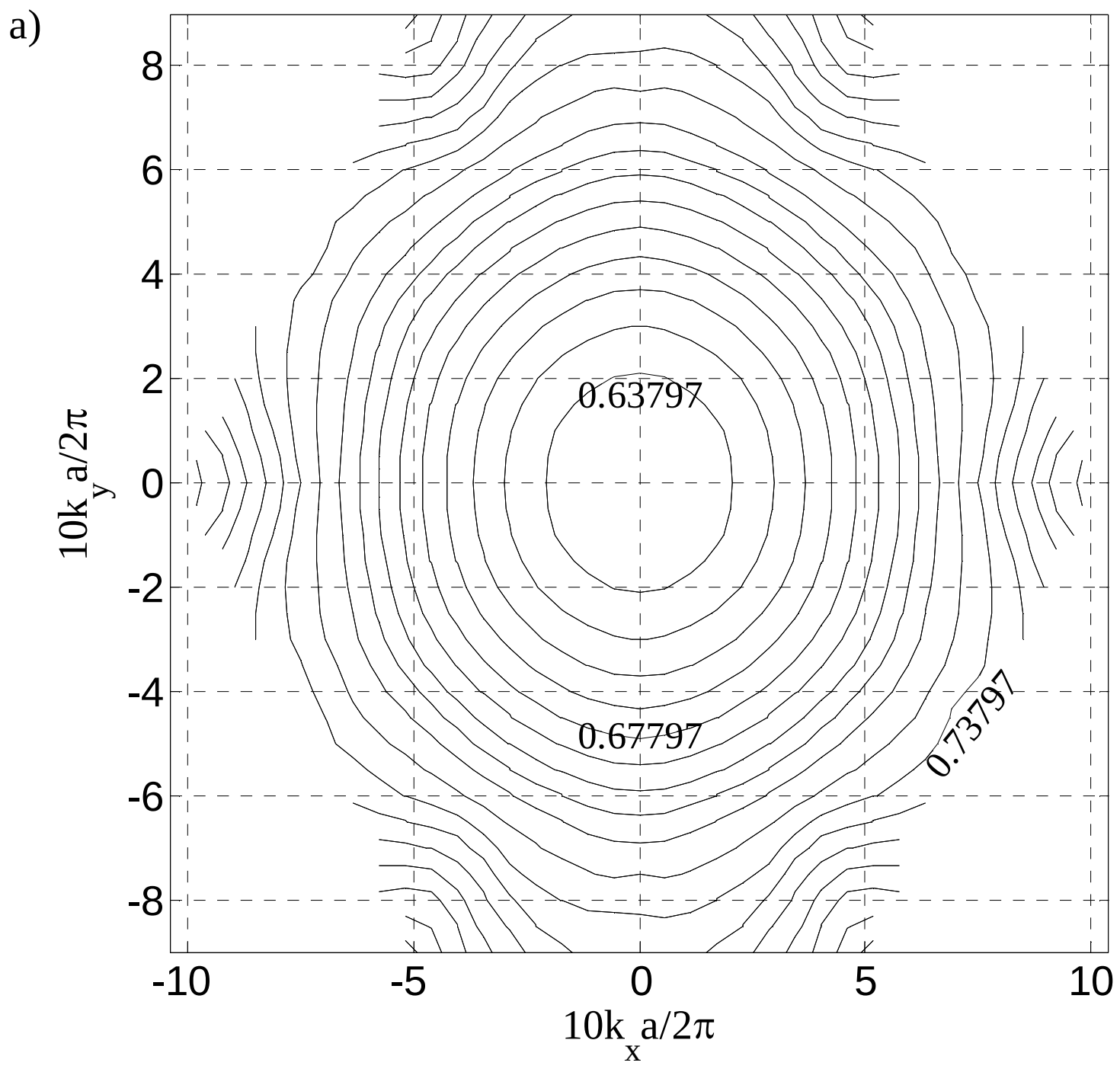

**Figure 6b**

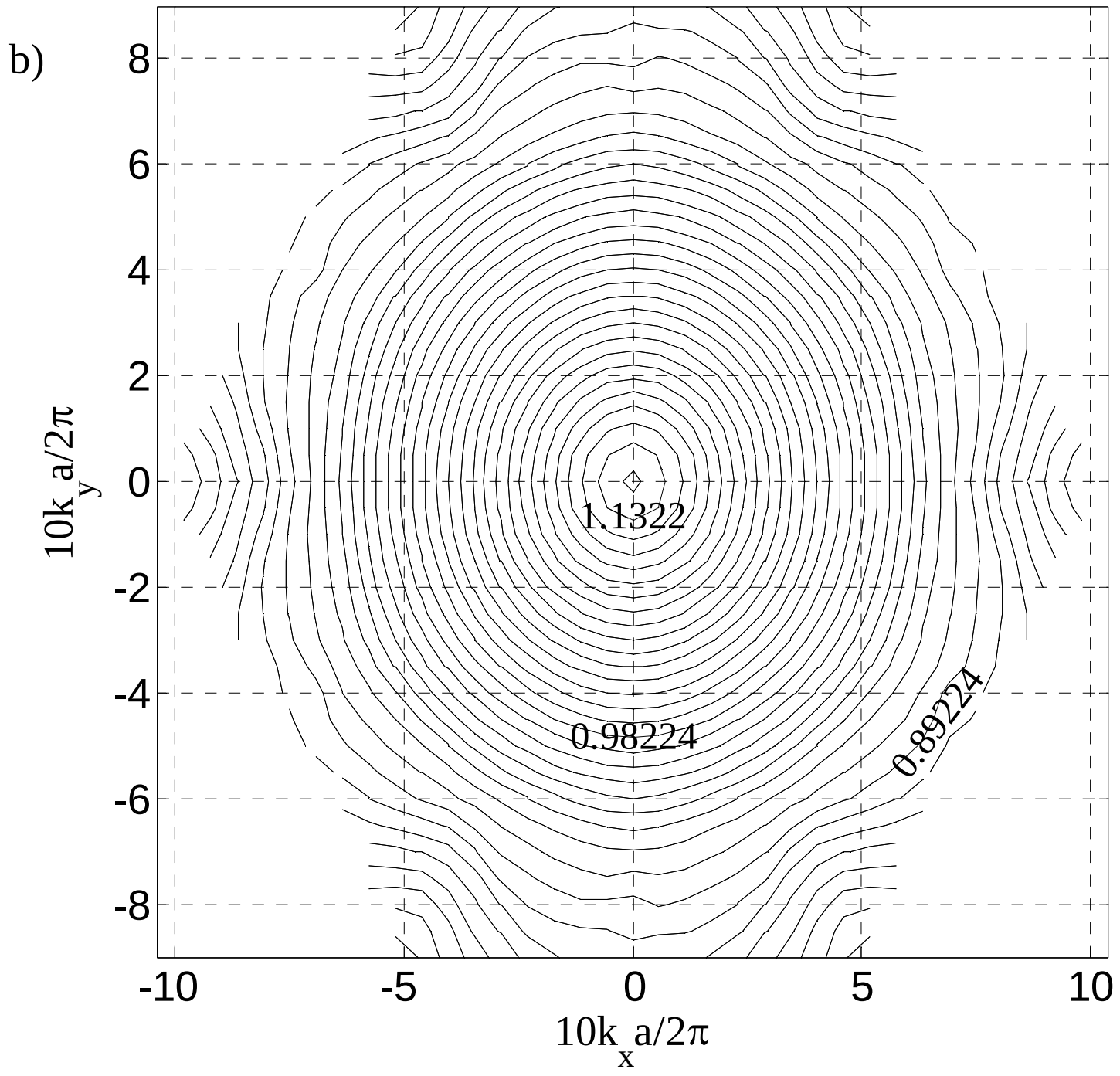

**Figure 6c**

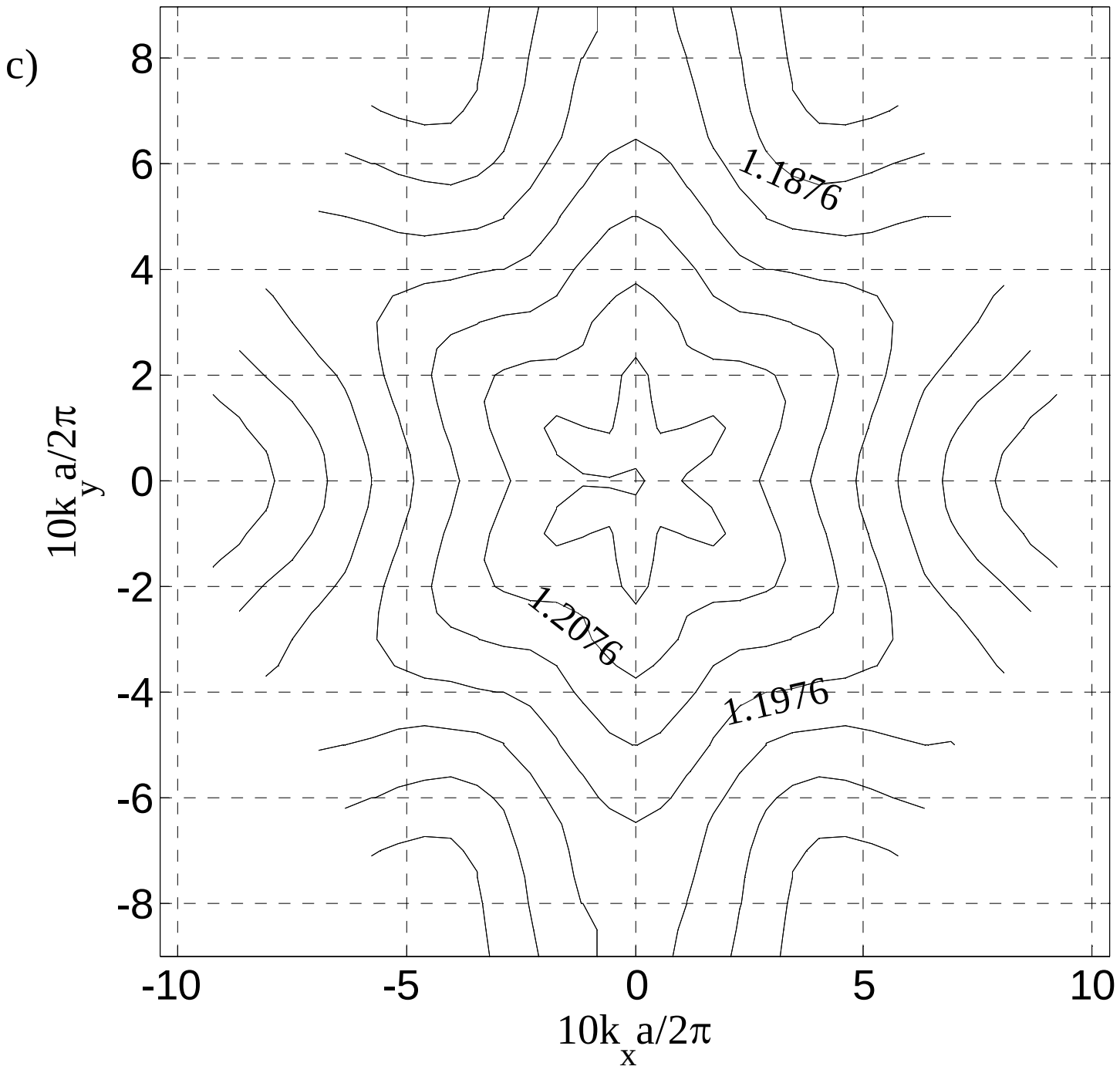

**Figure 7a**

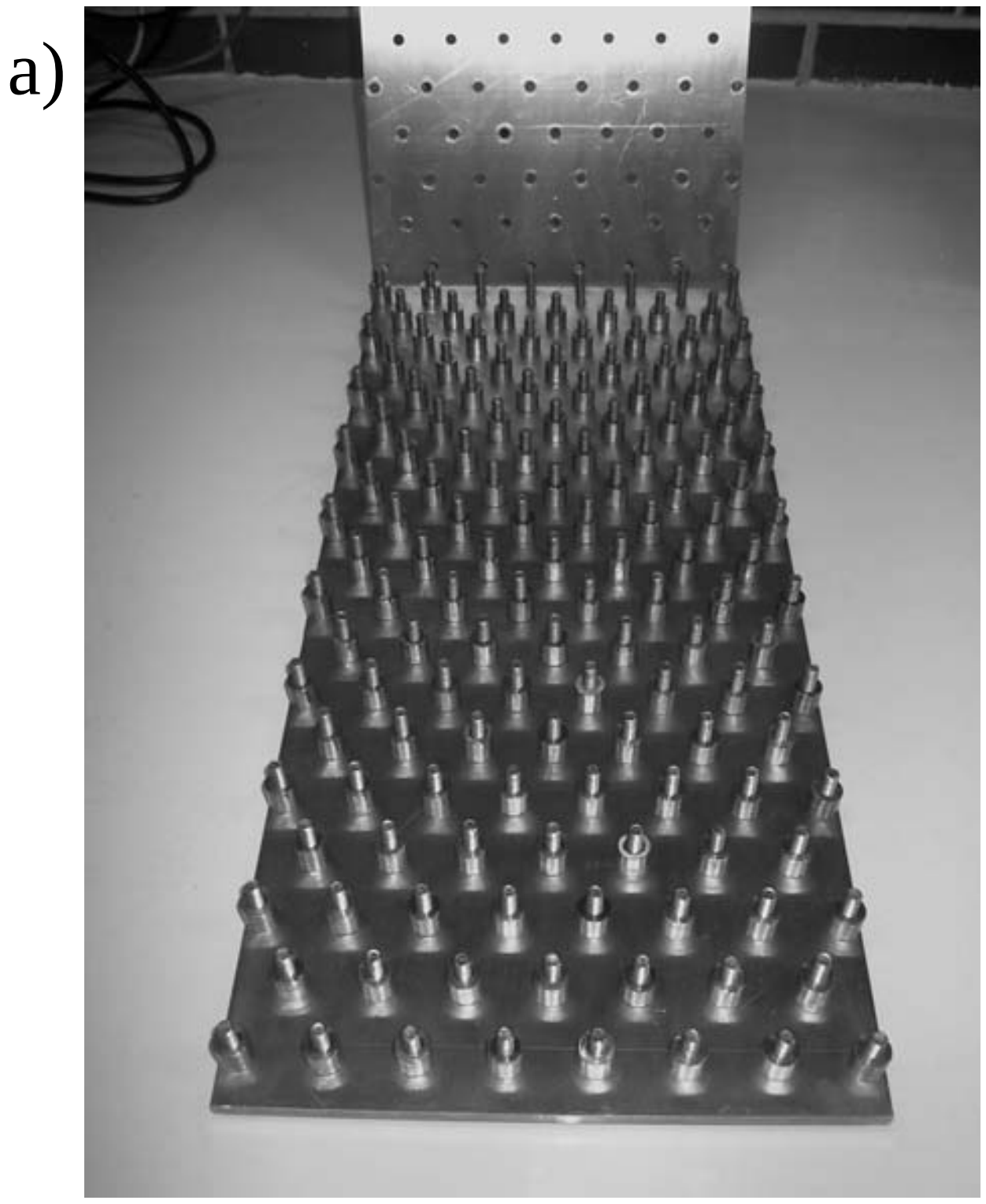

**Figure 7b**

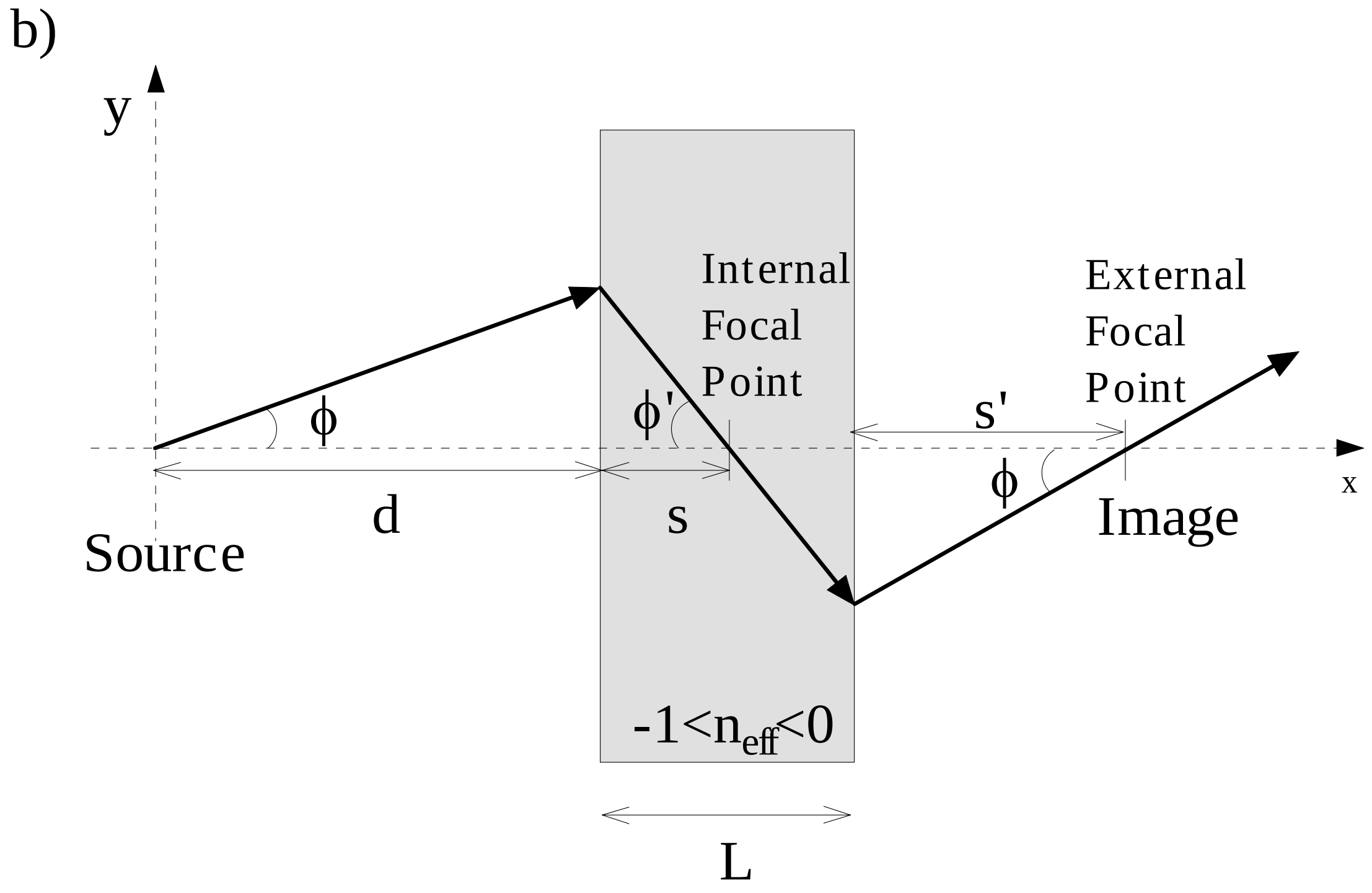

**Figure 8**

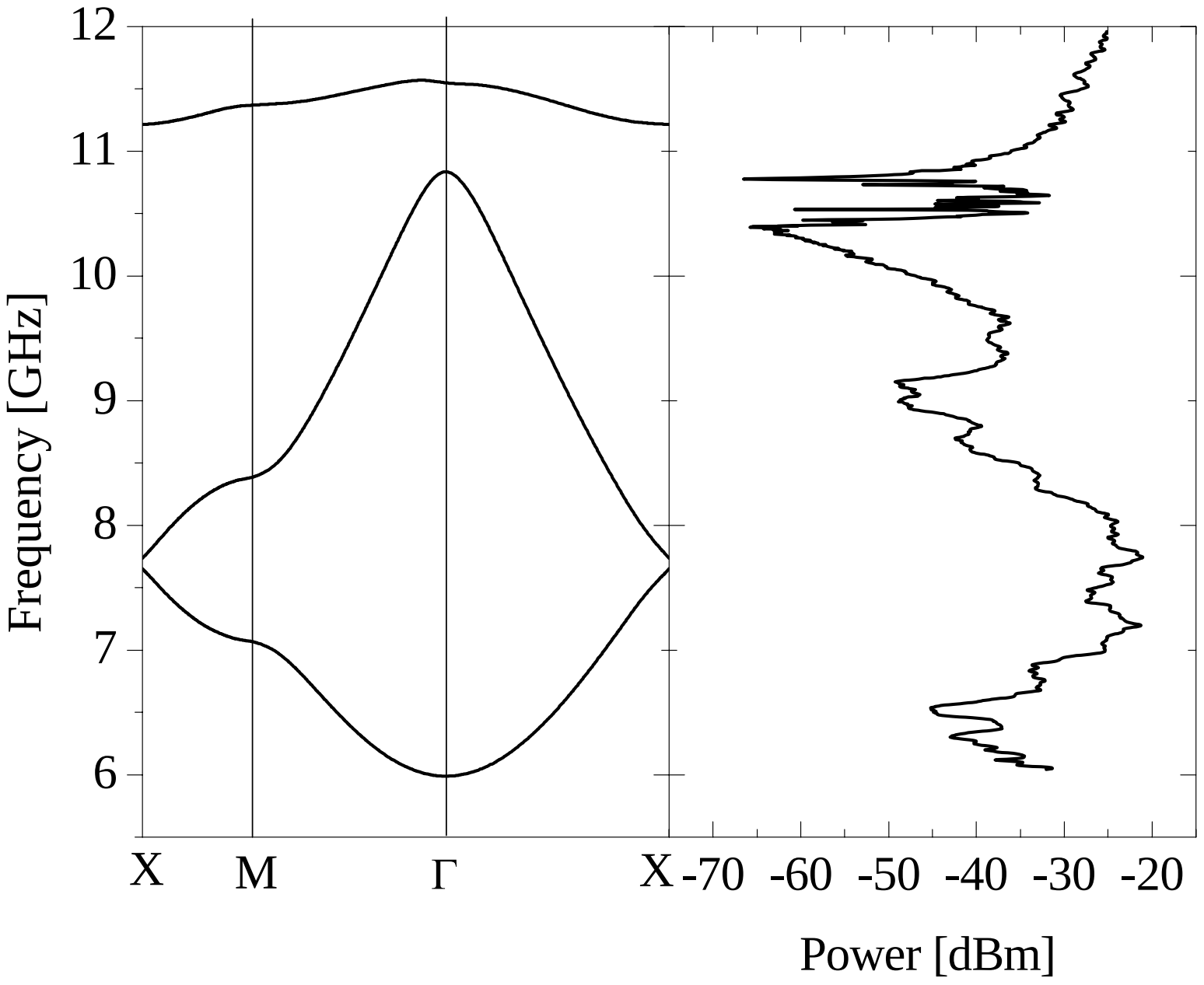

**Figure 9a**

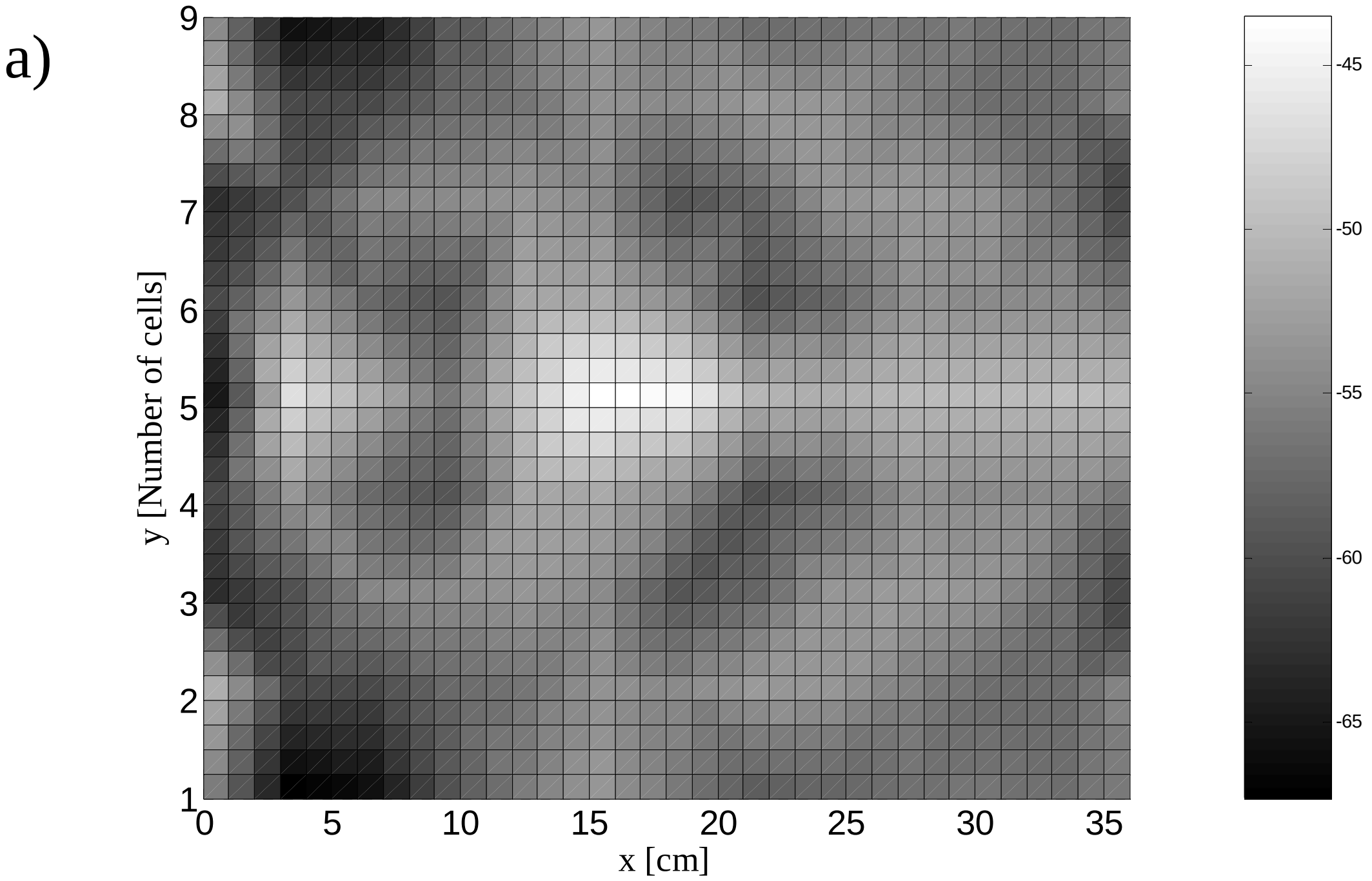

**Figure 9b**

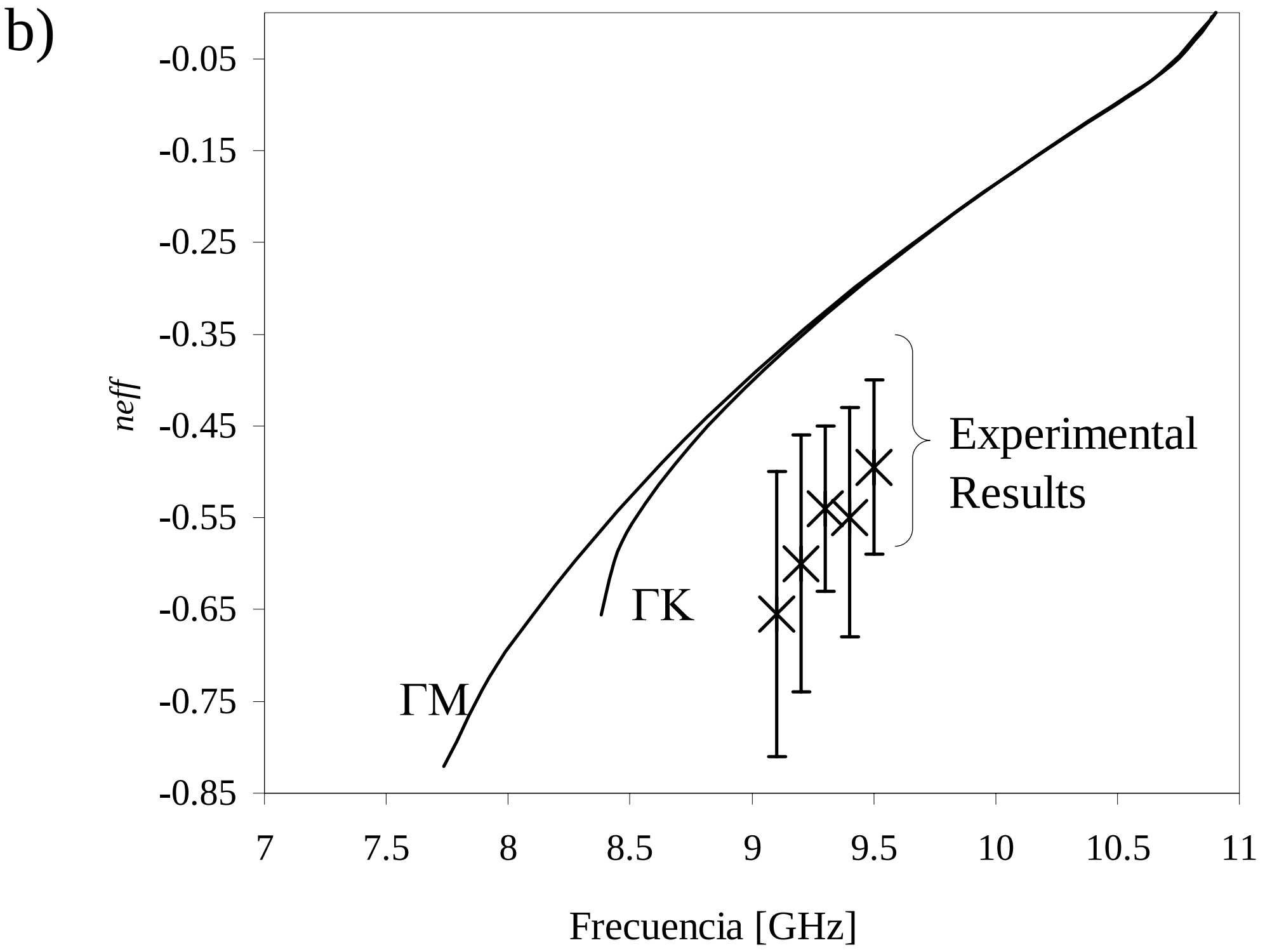

Figure 1: Scheme of two 2-D photonic crystals. The first one (a) is arranged into a square lattice and the second one (b) into a triangular lattice, both made out of metallic rods with a radius $r$ and a lattice constant $a$.

Figure 2: Reciprocal lattices and the first Brillouin zones for the square (a) and triangular (b) lattice.

Figure 3: Photonic band diagram for a metallic photonic crystal with a square (a) and triangular lattice (b), $r/a$=0.2.

Figure 4: Equi-frequency surface diagram for the first (a) and second (b) band of the 2-D metallic PC with a square lattice.

Figure 5: Effective refractive index against frequency for the TM modes of the first two bands of the PC with a square lattice (a) and a triangular lattice (b).

Figure 6: Equi-frequency surface diagram for the first (a), second (b) and third (c) band of the 2-D metallic PC with a triangular lattice.

Figure 7: Photograph of the built photonic crystal (a) and experimental scheme used to obtain the negative refractive index (b).

Figure 8: Experimental frequency response of the PC with a triangular lattice. Left panel: calculated band structure and right panel: measured transmission power.

Figure 9: (a) Typical mapping diagram for the autofocus effect taken at a frequency of 9.4 GHz and (b) summary graph showing the comparison between the experimental and simulated refraction index.